\documentclass[fleqn,usenatbib]{mnras}

\usepackage{newtxtext,newtxmath}
\usepackage[T1]{fontenc}
\usepackage{ae,aecompl}
\usepackage{xspace}
\usepackage{graphicx}	% Including figure files
\usepackage{amsmath}	% Advanced maths commands
\usepackage{amssymb}	% Extra maths symbols
\usepackage{xcolor}
\usepackage{fontawesome}

\usepackage{etoolbox}
\makeatletter
\patchcmd\@combinedblfloats{\box\@outputbox}{\unvbox\@outputbox}{}{%
   \errmessage{\noexpand\@combinedblfloats could not be patched}%
}%
 \makeatother

\newcommand{\Planck}{{\it Planck}\xspace}
\def\hmath$#1${\texorpdfstring{{\rmfamily\textit{#1}}}{#1}}
\title[P(k) from Ly$\alpha$ forest and others]{Matter power spectrum: from Ly$\alpha$ forest to CMB scales}

\author[S. Chabanier et al.]{Sol\`ene Chabanier$^{1}$\thanks{E-mail: solene.chabanier@cea.fr},
Marius Millea$^{2,3,4}$,
Nathalie Palanque-Delabrouille$^{1}$
\newauthor
\\\
$^{1}$ IRFU, CEA, Universit\'e Paris-Saclay, F91191 Gif-sur-Yvette, France\\
$^{2}$ Berkeley Center for Cosmological Physics, LBNL and University of California at Berkeley, Berkeley, California 94720, USA\\
$^{3}$ Institut d'Astrophysique de Paris (IAP), UMR 7095, CNRS-UPMC Universit\'e Paris 6, Sorbonne Universit\'es, 98bis boulevard Arago, F-75014 Paris, France\\
$^{4}$ Institut Lagrange de Paris (ILP), Sorbonne Universit\'es,
98bis boulevard Arago, F-75014 Paris, France}

\date{Accepted 2019 August 10. Received 2019 August 9; in original form 2019 May 22}

\pubyear{2015}

\begin{document}
\label{firstpage}
\pagerange{\pageref{firstpage}--\pageref{lastpage}}
\maketitle

% Abstract of the paper
\begin{abstract}
We present a new compilation of inferences of the linear 3D matter power spectrum at redshift $z\,{=}\,0$ from a variety of probes spanning several orders of magnitude in physical scale and in cosmic history. We develop a new lower-noise method for performing this inference from the latest Ly$\alpha$ forest 1D power spectrum data. We also include cosmic microwave background (CMB) temperature and polarization power spectra and lensing reconstruction data, the cosmic shear two-point correlation function, and the clustering of luminous red galaxies. We provide a Dockerized Jupyter notebook housing the fairly complex dependencies for producing the plot of these data, with the hope that groups in the future can help add to it. Overall, we find qualitative agreement between the independent measurements considered here and the standard $\Lambda$CDM cosmological model fit to the {\it Planck} data. 
\end{abstract}

% Select between one and six entries from the list of approved keywords.
% Don't make up new ones.

%%%%%%%%%%%%%%%%%%%%%%%%%%%%%%%%%%%%%%%%%%%%%%%%%%

%%%%%%%%%%%%%%%%% BODY OF PAPER %%%%%%%%%%%%%%%%%%
   \newcommand{\dd}{
      \mathop{}\mathopen{}\mathrm{d}
      }
\section {Introduction}

The $\Lambda$CDM model provides a simple and remarkable fit to much of the existing cosmological data, forming the basis of the standard cosmological paradigm. The cosmic microwave background (CMB) temperature and polarization anisotropies observed by the {\it Planck} satellite can be explained with only the six free parameters of the $\Lambda$CDM model \citep{PlanckCosmo2018, planck2018legacy}. In this paper, we illustrate the extent to which this model, with parameters fixed to their best-fit given {\it Planck} data, is in agreement with a number of other probes spanning cosmic time and cosmic scales. In an initial work, \cite{Tegmark2002} demonstrated the consistency between the $\Lambda$CDM model fit to the WMAP CMB data~\citep{WMAP9yr}, the first iteration of the Sloan Digital Sky Survey (SDSS~I) ~\citep{York2000} clustering data that were available at the time, the 2 Degree Field Galaxy redshift Survey(2dFGRS)~\citep{2dFGRS} galaxy clustering data and the Red-Sequence Cluster Survey~\citep{RSCS} weak lensing data. More recent updates to this work include \cite{tegmark2009} and \cite{hlozek2012}, which included newer data and other types of probes. With the advent of the Planck mission, of the third and fourth iterations of the Sloan Digital Sky Survey ~\citep{Blanton2017} and of the Dark Energy Survey \citep{DES}, the measurements have now reached an improvement of about an order of magnitude in precision over the last two decades since the initial work. These updated data sets make it timely to reevaluate the overall agreement.

\begin{figure*}	
\includegraphics[width=\textwidth]{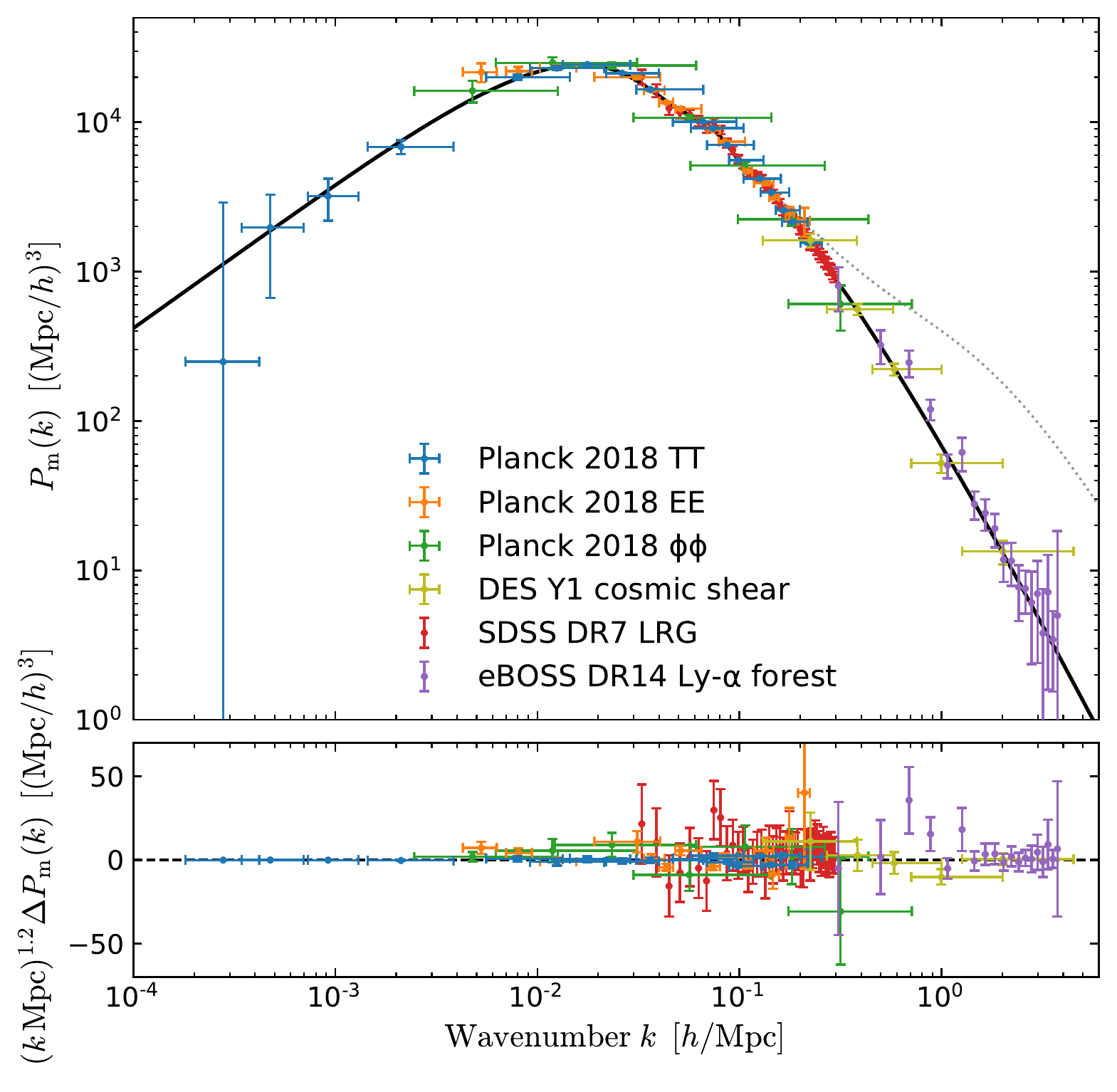}
\caption{{\it Top}: Data points show inferences of the 3D linear matter power spectrum at $z\,{=}\,0$ from Planck CMB data on the largest scales, SDSS galaxy clustering on intermediate scales, SDSS Ly$\alpha$ clustering and DES cosmic shear data on the smallest scales. In cases where error bars in the $k$-direction are present, we have used the method of \citet{Tegmark2002} to calculate a central 60\% quantile of the region to which each data point is sensitive. In other cases, data points represent the median value of the measurement. The solid black line is the theoretical expectation given the best-fit {\it Planck} 2018 $\Lambda$CDM model (this model also enters the computation of the data points themselves). The dotted line for reference shows the theoretical spectrum including non-linear effects. {\it Bottom}: deviation of the data from the Planck best fit $\Lambda$CDM 3D matter power spectrum.}
\label{fig:tegfig}
\end{figure*}

The main results of the paper are two-fold. First, focusing in particular on the Ly$\alpha$ constraints, we develop a new more accurate method for processing these data into a constraint on the linear matter power spectrum, $P_{\rm m}(k)$, at redshift zero. This method is based on a technique known as total variation regularization \citep[TVR;][]{Chartrand2005}, which reduces noise in the resulting estimate. Second, we take this constraint, combined with a number of others, and produce a compilation of $P_{\rm m}(k)$, shown in Fig.~\ref{fig:tegfig}. On scales of a few Mpc, we include the information embedded in the  Ly$\alpha$ forest measured with the quasar survey of the SDSS~IV fourteenth data release \citep{DR14eB}. Partially overlapping in scale, we also use the cosmic shear measurement from the DES YR1 data release \citep{troxel2017}. On scales of several tens of Mpc, we use the power spectrum of the  halo density field derived from a sample of luminous red galaxies (LRG) from the SDSS seventh data release (DR7) \citep{reid2010a}. Finally, on the largest scales, we use the anisotropies of the microwave background measured by the Planck satellite. In addition to probing a wide range of scales, from $k=2\times 10^{-4}$ to $k=2\,h\,\rm Mpc^{-1}$, these data also cover a large range of cosmic epochs: $z\sim 0.35$ for the LRG, $z\sim 0.2$ to 1.3 for the shear measurements, $z= 2.2$ to 4.6 for the more distant Ly$\alpha$ forest, and $z\sim 10^3$ for CMB.

As described in \cite{Tegmark2002}, inferring the linear matter spectrum at $z\,{=}\,0$ from the various probes we consider here is a highly model-dependent process. We take as our fiducial model the Planck 2018 best-fit $\Lambda$CDM model \citep{PlanckCosmo2018}. The results here are therefore a test of the consistency of this model, rather than direct constraints on the matter power spectrum. In general, we find qualitative agreement of this fiducial model with the data we consider.

The datasets which we consider were chosen to be representative of different types of cosmological measurements which exist and to cover a broad range of scales, particularly favoring ones where data products were especially convenient for the calculations we perform here. Of course, many other measurements exist which provide constraints on the matter power spectrum, some of which are known to be in varying degrees of tension with the {\it Planck} best-fit model. It is beyond the scope of this work to include them all, however we provide a Dockerized Jupyter notebook which includes the fairly complex dependencies needed to produce this plot. We hope that this makes it easy for any group in the future to add any desired data set and keep up-to-date this compilation. The repository for this notebook can be found here: \href{https://github.com/marius311/mpk_compilation}{\faGithub}\footnote{\url{https://github.com/marius311/mpk_compilation}}.

The outline of the paper is as follows. In Sec.~\ref{sec:Lya}, we present the Ly$\alpha$ data and explain how we compute the 3D matter power spectrum from the published 1D flux power spectrum. These data  are  the ones whose treatment differs the most from the previous study of \cite{Croft2002} used in~\cite{Tegmark2002}. In Sec.~\ref{sec:oprobes}, we present the other probes we use (CMB, cosmic shear and galaxy clustering) and the general method we apply to compute the 3D matter power spectrum in each case. We conclude in Sec.~\ref{sec:conclusions}.

\section{Matter power spectrum from the Lyman-alpha forest}
\label{sec:Lya}
\subsection{Lyman-alpha data}
With the advent of medium-resolution spectroscopic surveys that increased by several orders of magnitude the number of spectroscopically observed high-redshift quasars, the past decade has witnessed a significant ramp up of the use of Ly$\alpha$ forest as a cosmological probe. The Sloan Digital Sky Survey in particular, with the BOSS and eBOSS surveys~\citep{Dawson2016,Blanton2017}, has now observed over two hundred thousand quasar spectra at redshifts above 2.1. The 3D correlations in the Ly$\alpha$ flux transmission field were studied extensively in \cite{Slosar2011,Busca2012,Slosar2013,Kirkby2013,Delubac2015,Bautista2017} to measure the position of the Baryon Acoustic Oscillation peak and provide constraints on dark matter and dark energy. The correlations along the line of sight provide information on smaller scales. The 1D flux power spectrum measured from the Ly$\alpha$ data, for instance, is a remarkable probe of the impact on structure formation of neutrino masses~\citep{Palanque2015a,Palanque2015b,Yeche2017}, of warm dark matter \citep{Baur2016,Irsic2017, Armengaud2017} or of various models of sterile neutrinos~\citep{Baur2016,Baur2017}.

In this work, we use data from the eBOSS DR14 release~\citep{DR14eB}, corresponding to the entirety of the BOSS survey complemented by the first year of eBOSS. We take the 1D transmitted flux power spectrum measured by the BOSS and eBOSS collaborations in \cite{Chabanier2018}. It used a selection of 43,751 highest-quality quasar spectra from a parent sample of 180,413 visually inspected spectra. They were selected for the absence of BAL features, a good mean spectral resolution and high mean signal-to-noise ratio per pixel. The analysis gives the flux power spectrum along a line of sight, $P_{\rm tot\,1D}$, in thirteen equally-spaced redshift bins covering the range $z=2.2$ to 4.6 with $\Delta z=0.2$. The highest redshift bin is built from 63 quasars only and has large uncertainties. We therefore use only the lowest twelve redshift bins in this work. These data show an oscillatory feature due to the correlated absorption by Ly$\alpha$ and SiIII at a velocity separation  $\Delta v = 2271~\rm km.s^{-1}$. Adopting the approach from \cite{McDonald2006}, we model the total transmission flux fraction as,
\begin{equation}
\delta _ { \mathrm { tot } } = \delta ( v ) + a \delta ( v + \Delta v ),
\end{equation}
with $\delta ( v )$ being only for Ly$\alpha$. The resulting power spectrum is 
\begin{equation}
\label{eq:si3}
P_{\rm tot\, 1D}(k) = (1+a^2) P_{F1D}(k) + 2a\cos(\Delta v\,k)P_{F1D}(k)
\end{equation}
We use equation \ref{eq:si3} to correct for these wiggles, where $a$ is fit independently for each redshift bin. 
We use these 1D transmitted flux power spectra to derive the 3D matter power spectrum as explained below.

\subsection{Method}

We follow the prescription of \cite{Croft1998}, updated in~\cite{Croft2002}. We assume that the 3D flux power spectrum $P_{F3D}$ is  related to the linear matter power spectrum $P_{{\rm m}}$ by a proportionality relation,
\begin{equation}
    P_{{\rm m}}(k, z)= \frac{P_{F3D}(k, z)}{b^2(k, z)}\;,
	\label{eq:bias}
\end{equation}
with $b(k,z)$ a scale and redshift dependent bias that depends on the cosmological model. The scale dependence is an improvement over the initial methodology, added in \cite{Croft2002}, to take into account the effects of non-linear evolution, thermal broadening and peculiar velocities. 

The 1D and the 3D flux power spectra are related by
\begin{equation}
    P_{F3D}(k) = -\frac{2\pi}{k}\frac{\dd P_{F1D}(k)}{\dd k}\;,
	\label{eq:rel1d3d}
\end{equation}
which we use to derive  the 3D flux power spectrum needed in Eq.~\eqref{eq:bias}.

We compute the bias $b(k,z)$ for each of the twelve redshift bins mentioned above using CAMB\footnote{\url{https://camb.info} \citep{Lewis2000} for the linear matter power spectrum, and hydrodynamic simulations dedicated to the analysis of the BOSS 1D data~\citep{Borde2014} for the 1D flux power spectrum}. The simulations are produced for a grid of parameters whose values are varied around a central model. The four cosmological parameters are the scalar spectral index $n_s$, the RMS matter fluctuations amplitude  today in linear theory $\sigma_8$,  the matter density today $\Omega_m$, and the expansion rate today $H_0$. The astrophysical parameters (all at $z=3$) are  the normalization temperature of IGM $T_0$, the logarithmic slope of the $\delta$ dependence of the IGM temperature $\gamma$,  the  effective optical depth of the Ly$\alpha$ absorption $A^\tau$ and  the logarithmic slope $\eta^\tau$ of the redshift dependence of $A^\tau$. The central (also dubbed best-guess) simulation is based upon a fiducial model corresponding to the \cite{PlanckCollaboration2013} best-fit cosmology. The simulation grid, however, allows us to test other cosmologies.  

\begin{figure}	\includegraphics[width=\columnwidth]{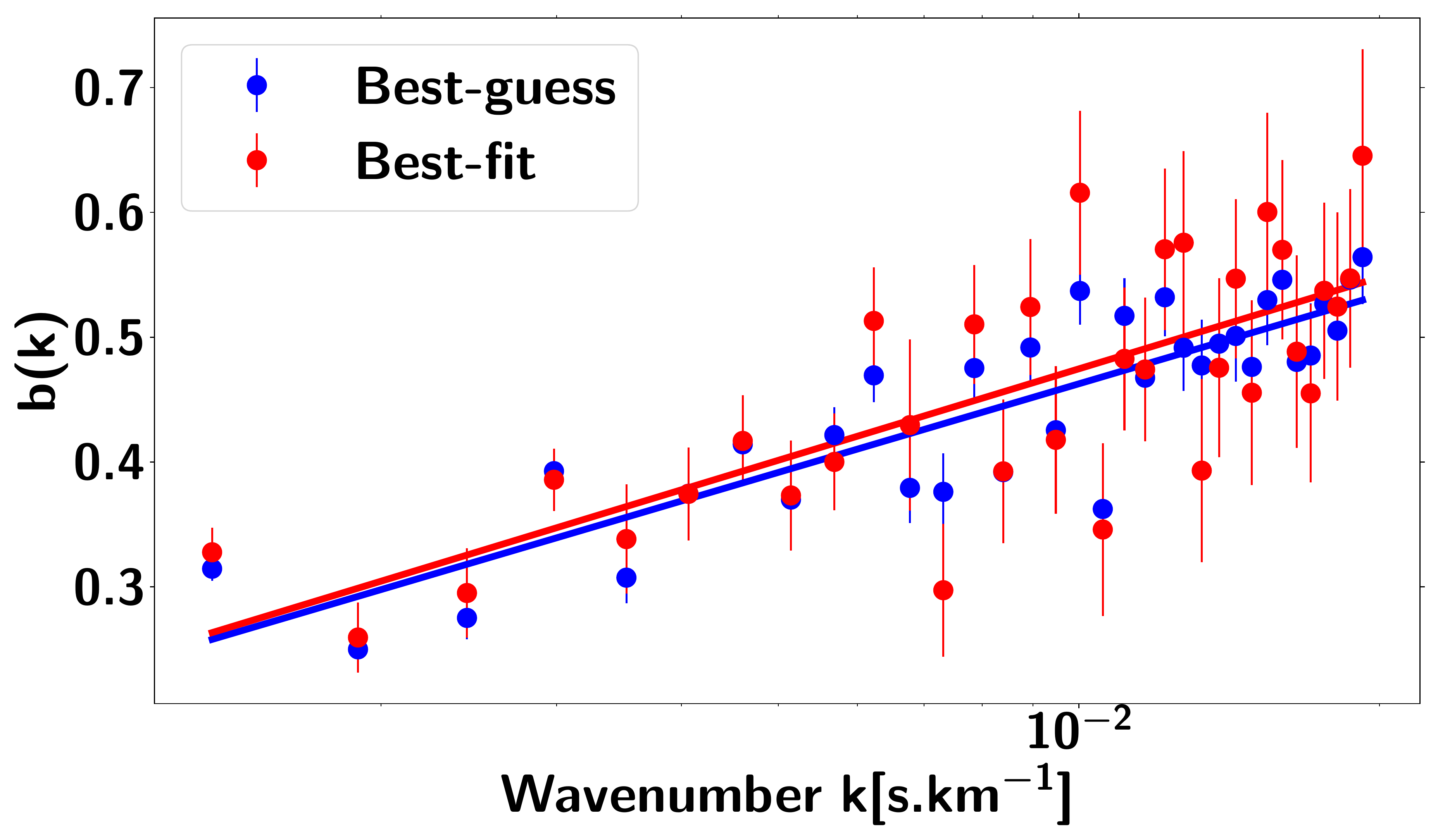}
    \caption{Biases computed at $z=2.8$ for the best-guess (in blue) and  best-fit (in red) configurations. The lines are linear-log fits to each case.}
    \label{fig:biases}
\end{figure}

In Table \ref{tab:bgbf}, we list  the  values of the  parameters used in the best-guess  simulation, as well as the corresponding best-fit values measured  in \cite{Chabanier2018}, for a fit to the eBOSS 1D Ly$\alpha$ power spectrum combined with the Planck 2018 ``TT+lowE" likelihood~\citep{PlanckCosmo2018}. The best-fit model  is in good agreement with the central simulation.
The parameters that deviate the most from their central value are $\sigma_8$ and $\Omega_m$. 
We determine the biases $b_{\rm bf}$ for the best-fit model by computing the biases $b_{\rm bg}$ for the best-guess simulation, and we apply first-order corrections to account for the measured shifts in $\sigma_8$ and $\Omega_m$, using simulations where all parameters are kept to their central value except for either $\sigma_8$ or $\Omega_m$. 
\begin{table}
	\centering
	\caption{Fit parameters. First column: central value and variation range in the simulation grid. Second  column: best-fit value and 68\% confidence interval for a fit to Ly$\alpha$ + Planck (TT + lowE). }
	\label{tab:bgbf}
	\begin{tabular}{ccc}
		\hline
		 Parameter& Simulations  & Best-fit \\ \hline
         $n_s$ & $0.96 \pm 0.05$ & $0.954 \pm 0.004$ \\
         $\sigma_8$ & $0.83 \pm 0.05$ & $0.817 \pm 0.007$\\
         $\Omega_m$ & $0.31 \pm 0.05$ & $0.330 \pm 0.009$\\
         $H_0 \;(km.s^{-1}.Mpc^{-1})$ & $67.5 \pm 5$& $66.2 \pm 0.6$  \\
        \hline
        $T_0(z=3) \;(K)$ & $14 000 \pm 7 000$ & $11 300 \pm 1 600$\\
        $\gamma(z=3)$ & $1.3 \pm 0.3$ & $0.7 \pm 0.1$ \\
        $A^\tau$ & $0.0025 \pm 0.0020$ & $0.0026\pm 0.0001$\\
        $\eta^\tau$ & $3.7\pm 0.4$& $3.734 \pm 0.015$\\
        \hline
	\end{tabular}
\end{table}
We determine the bias $b(z,k)$ at each redshift $z$ and scale $k$  by 
\begin{align*}
    b_{\rm bf}(z,k) &= b_{\rm bg}(z,k)\\
                           &+ (\sigma_{8, \rm bf}-\sigma_{8, \rm bg})\frac{\dd b}{\dd \sigma_8}(\sigma_{8,\rm bg},\Omega_{m,\rm bg})\\
                            &+ (\Omega_{m,\rm bf}-\Omega_{m, \rm bg})\frac{\dd b}{\dd \Omega_m}(\sigma_{8,\rm bg},\Omega_{m, \rm bg})\;.
	\label{eq:teb}
\end{align*}
Fig.~\ref{fig:biases} shows both best-guess and best-fit biases for redshift  $z=2.8$. As illustrated in the figure for a specific redshift, but similarly for all redshifts, the linear corrections have little effect.

Eq.~\eqref{eq:bias} thus allows us compute the linear power spectra $P_{{\rm m}}(k,z_{i})$ for all twelve redshift bins  $z_{i}$.  We then transpose each of them to $z=0$ with the relation
\begin{equation}
    P_{{\rm m},z_{i}}(k, 0)=P_{\rm m}(k,z_{i})\times t(k,z_{i})\;,
	\label{eq:tpm}
\end{equation}
where the evolution term $t(k,z_i)$ is determined in linear theory using a Boltzmann code such as  CAMB\footnote{\url{https://camb.info}} \citep{Lewis2000} or  \textsf{CLASS}\footnote{\url{http://class-code.net/}} \citep{CLASS}. Finally we combine all twelve $z=0$ power spectra $P_{{\rm m},z_{i}}$ using an inverse-variance weighted average. The top panel of Fig.~\ref{fig:pm0lya} shows the resulting $P_{{\rm m}}(k,0)$.

\subsection{Total Variation Regularization}
\begin{figure*}
	\includegraphics[width=\textwidth]{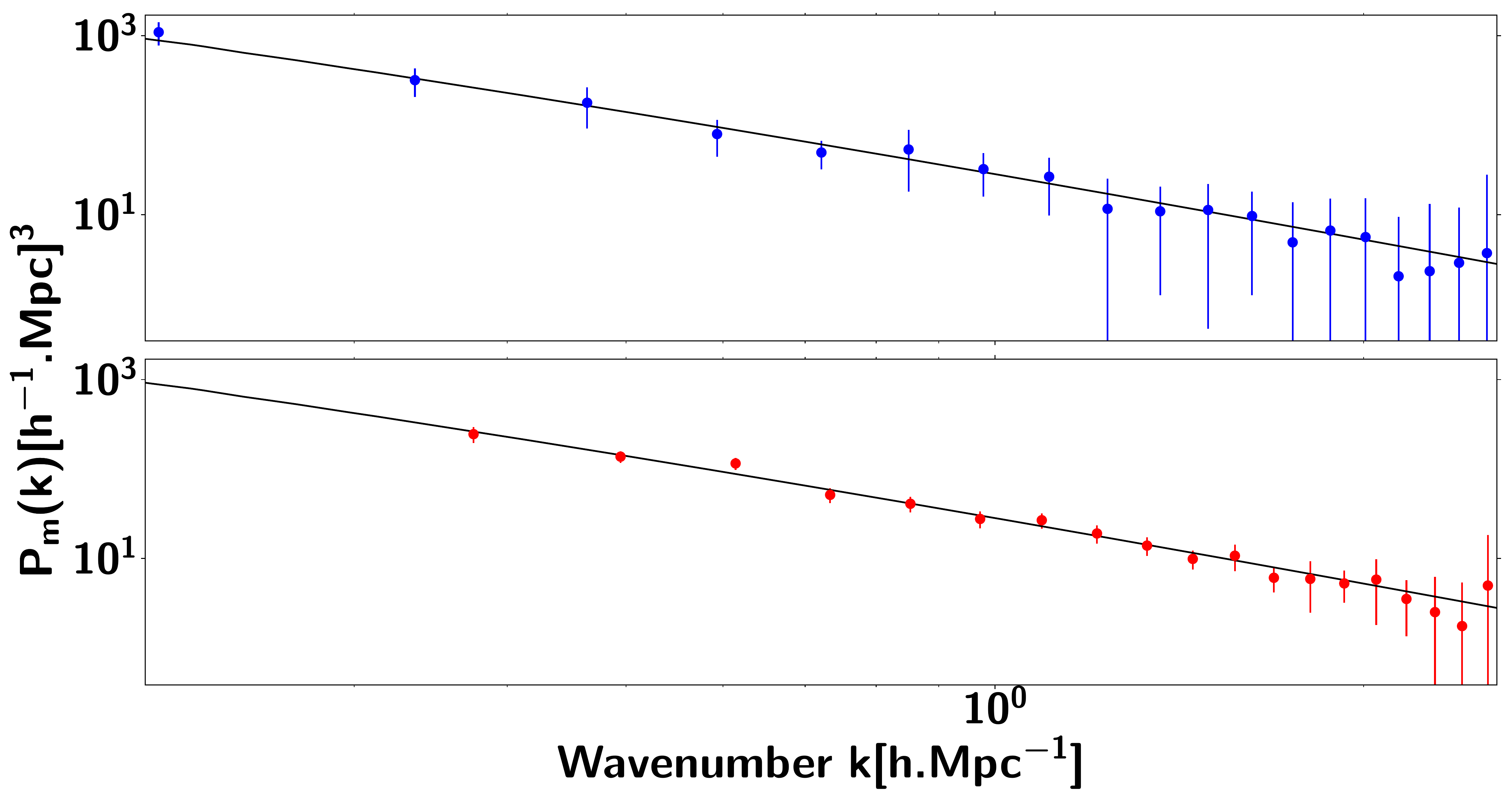}
    \caption{Linear matter power spectrum inferred from Ly$\alpha$ data. Results from the discrete differentiation are shown in the top panel, and from the TVR approach in the bottom panel. The black solid line is the linear theory expectation.}
    \label{fig:pm0lya}
\end{figure*}

The discrete differentiation of the 1D transmitted flux power spectrum $P_{F1D}$ to obtain the 3D transmitted flux power spectrum $P_{F3D}$ significantly amplifies noise and uncertainties. The effect is worst at small scales where only the highest redshift bins, which are also the noisiest,  contribute to the measurement. To reduce this computational artifact, we use a refined differentiating technique, the total variation regularization (TVR) method, proposed in \cite{Chartrand2005}. It is a specific regularization process that estimates the derivative of a function $f$ as the minimizer $u_{min}$ of the functional F,
\begin{equation}
F ( u ) = \alpha R ( u ) + DF ( A u - f ),
\end{equation}
where $\alpha$ is the regularization parameter, $R(u)$ is the regularization term which penalizes noise, and $DF ( A u - f )$ is the data fidelity term with $Au ( x ) = \int _ { 0 } ^ { x } u$. The TVR uses $R(u) = \int  \left| u ^ { \prime } \right|$ and $D F ( \cdot ) = \int  | \cdot | ^ { 2 }$. The resulting algorithm has only one free parameter, $\alpha$, that we fix to $10^{-5}$ for all the redshift bins, as it appears to be a good compromise between smoothing the data and conserving valuable information. We tested the TVR on an analytical form of the 1D flux power spectrum, which allowed us to compare the resulting derivative to the true $P_{F3D}$. The  TVR induces no computational bias, except on the first three sampling points, which we hence decide not to keep in the following.
 To estimate the uncertainty on the 3D power spectrum resulting from this regularization, we perform a parametric bootstrap at each $k$ bin with 1000 iterations. 
The bottom panel of Fig.~\ref{fig:pm0lya} shows the final 3D matter power spectrum at redshift $z = 0$ derived  with the TVR approach. The dispersion is clearly reduced and the power spectrum from TVR considerably smoother than the one from a straight derivative.  The TVR technique increases the  correlations between neighboring points (up to 50\% in the worst case, for nearest-neighbor correlation), although correlations with  next-to-nearest neighbors are between 1 and 20\% at most.

Finally, we point out that we use the TVR derivation  for the data but we keep to straight derivatives to compute   the biasing functions from the hydrodynamic simulations. The reason is the  following. The $P_{F1D}$ from the simulations is much smoother than in the data, and systematic uncertainties from the bias term are largely sub-dominant compared to data  statistical uncertainties. Using the TVR technique on the simulations would therefore unnecessarily   increase the correlations between neighboring points without yielding a measurable gain on the resulting uncertainties.

\section{Matter power spectrum from other probes}
\label{sec:oprobes}

Having described in some detail the Ly$\alpha$ forest constraints and our new TVR-based method for calculating them, we now turn to constraints from the other datasets considered in this work, which more closely follow the procedure laid out by \cite{Tegmark2002}. Their procedure is based on the relating a given observable, $d_i$ (which can be for example a CMB $C_\ell$, or measurement of cosmic shear power spectrum at some redshift, etc...), to $P_{\rm m}(k,0)$, via
\begin{align}
d_i = \int d \ln k \, W_i(k) P_{\rm m}(k,0) 
\label{eq:canonical}
\end{align}
Each given observable will have a different ``window'' function, $W_i(k)$, which can be calculated from theory for a fixed cosmological model. In many cases, for example if our $d_i$ are simple auto-correlation functions, the $W_i(k)$ are strictly positive. Furthermore, depending on the exact quantity measured, they are often also fairly localized in $k$. In these cases, we normalize the $W_i(k)$ to unit area, effectively treating it as a probability distribution, and, following \cite{Tegmark2002}, take the error bar in the $k$-direction in Fig.~\ref{fig:tegfig} to denote the middle 80\% quantile of this distribution. Our slight modification to their procedure is that whereas they take the middle 80\% of the quantity $W_i(k) P_{\rm m}(k,0)$, we take it of just $W_i(k)$. We view this as the more natural choice since it is just $W_i(k)$ which represents the projection of the data into the redshift zero matter power spectrum. Additionally, this gives us a $k$-direction error bar which does not depend on the shape of $P_{\rm m}(k,0)$. 

\begin{figure}	
    \includegraphics[width=\columnwidth]{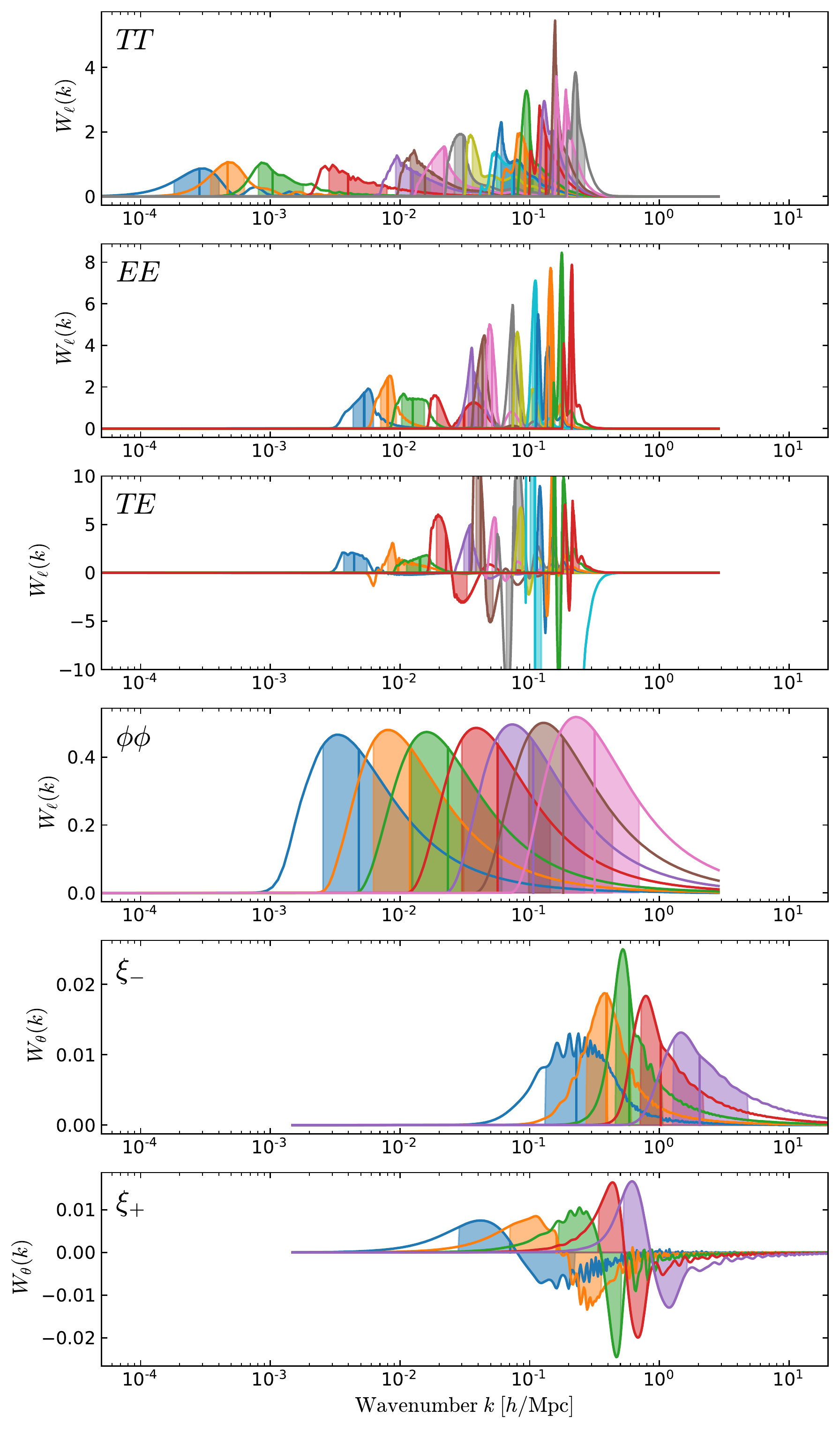}
    \caption{The window functions, $W_i(k)$, for several of the datasets considered here. The shaded region represents the middle 80\% quantile of the absolute value of the function, which is the region denoted by the $k$-direction error bars in Fig.~\ref{fig:tegfig}, and represents roughly to which $k$-scales a given data observation is sensitive to. Note that some observations have non-strictly positive windows, meaning we cannot interpret them as simply a measure of the overall amplitude of the matter power spectrum at a given scale, but rather some combination of this and its derivative.}
    \label{fig:windows}
\end{figure}

In Fig.~\ref{fig:windows}, we plot the window functions for the different observations which we use. In each case, some ``rebinning'' of the data is applied as compared to the raw data products provided by each experiment. This is done so as to produce more reasonably spaced data points in the $k$ direction, and to improve the localization of the $W_i(k)$. We describe these rebinnings in the individual sections below. One can verify the localized nature of the different window functions, indicating the validity of interpreting each data point as a constraint on $P_{\rm m}(k,0)$. 

\

\noindent {\bf Cosmic microwave background} \; For CMB data, we use the \Planck 2018 temperature, polarization, and lensing reconstruction power spectra \citep{planck2018legacy,planck2018lensing}. At $\ell\,{<}\,30$ in temperature, we use the $C_\ell$'s provided by the \textsc{Commander} likelihood, with the asymmetric errorbars averaged together, which should have minimal impact as we also bin multiple $C_\ell$'s together which will have a symmetrizing effect. At $\ell\,{>}\,30$ in temperature and polarization, we use the \textsc{Plik-like} bandpowers and covariance, rebinned as described above. We do not use polarization below $\ell\,{<}\,30$ because the signal there is highly reionization-model dependent \citep[e.g.,][]{zaldarriaga1997a}. For the lensing reconstruction, we use the bandpowers and covariance from the ``agressive'' data cut. The window functions are shown in Fig.~\ref{fig:windows}. One can see that the TE window functions are not strictly positive since they do not arise from an auto spectrum. For this reason, we cannot interpret them as a constraint on the amplitude of $P_{\rm m}(k,0)$, hence we show only TT and EE in Fig.~\ref{fig:tegfig}. Although we do not do so here, one could interpret them as a constraint on a linear combination of the amplitude and derivative of $P_{\rm m}(k,0)$, however. 

\

\noindent {\bf Cosmic shear} \; For cosmic shear, we use DES first-year constraints on the cosmic shear real-space two-point correlation functions $\xi_\pm^{ij}(\theta)$, where the $i$ and $j$ indices label different redshift bins \citep{troxel2017}. These functions can be written in the form of Eq.~\eqref{eq:canonical},
\begin{align}
\xi_\pm^{ij}(\theta) =\int d\ln k\; W^{ij}_\pm(\theta,k) P(k,0),
\end{align}
where
\begin{align}
W^{ij}_\pm(\theta,k) = \frac{1}{2\pi} \int_0^{\chi^H} d\chi \; \ell(\ell+1/2) J_{0/4}(\theta\ell)  \frac{q^i(\chi)q^j(\chi)}{\chi^2} \frac{P(k,\chi)}{P(k,0)},
\end{align}
the $q^i(\chi)$ are the lensing efficiency functions defined as usual \citep[e.g. as in][]{troxel2017}, and
\begin{align}
k = \frac{\ell+1/2}{\chi}.
\end{align}
We choose to bin together all of the redshift bins, producing a set of 5 fairly localized window functions for each $\theta$ bin, plotted in Fig.~\ref{fig:windows}. Interestingly, one can see that $\xi_+$ produces window functions which are not strictly positive. This arises due to the weighting of the Bessel function inside of the integrand. Thus, similarly as for the CMB TE power spectrum, we do not plot these constraints on Fig.~\ref{fig:tegfig}, although they could in theory also be interpreted as a joint constraint on the amplitude and derivative. 

\

\noindent {\bf Galaxy clustering} \; For galaxy clustering, we use measurements of the halo power spectrum from a sample of luminous red galaxies from the Sloan Digital Sky Survey seventh data release \citep{reid2010a}. Using a model for the halo bias, we can relate these measurements to the underlying linear matter power spectrum in which we are interested. We use the model given in \cite{reid2010a} with free parameters $b_0, a_1$, and $a_2$. Fitting to our fiducial cosmological model, we find best-fit values of $1.24, 0.54$, and $-0.33$, respectively, at a pivot scale of $k_\star=0.2\, {\rm Mpc}/h$.

\section{Conclusions}
\label{sec:conclusions}
In this letter, we present a measurement of the 3D matter power spectrum at redshift $z=0$ by combining different cosmological probes spanning four orders of magnitude in scales, from $k=2\times 10^{-4}$ to $k=2\,h\,\rm Mpc^{-1}$, and  a wide range of cosmic history, from $z\sim 0$ to 1000, shown in Fig.~\ref{fig:tegfig}. Taking advantage of the advent of new generation instruments to probe cosmic structure, we re-evaluate the study done in \cite{Tegmark2002}.  We use the latest data sets available for the  Ly$\alpha$ forest 1D power spectrum (SDSS-IV DR14), for the Cosmic Microwave Background temperature anisotropies and polarization measurements (Planck 2018), for the cosmic shear two-point correlation function (DES YR1) and for the galaxy clustering with measurements of the halo power spectrum (SDSS DR7).

On scales of a few Mpc we use measurements of the 1D transmitted flux Ly$\alpha$ power spectrum measured by the BOSS and eBOSS surveys of the SDSS \citep{Chabanier2018}. We follow the general method of \cite{Croft2002} to recover the 3D matter power spectrum from the 1D measurements. 
However, we improve the determination of the total power  $P_{\rm m}(k,0)$ compared to the previous analysis by using a refined differentiation technique, the Total Variation Regularization method proposed in \cite{Chartrand2005}, which allows us to significantly reduce the resulting dispersion. 
On small scales, we also use cosmic shear real-space two-point correlation functions from the DES YR1 data release \citep{troxel2017}, which undergo the same general treatment as in \cite{Tegmark2002}.
On scales of tens of Mpc, we use measurements of the halo power spectrum from a sample of LRGs  from the SDSS seventh data release. We use the halo bias model from \cite{reid2010a}.
For scales of hundreds of Mpc we use CMB data with temperature, polarization and lensing reconstruction power spectra measurements  \citep{planck2018legacy,planck2018lensing}. Except for Ly$\alpha$ that undergoes a specific treatment, we  apply the general method of \cite{Tegmark2002} to estimate the amplitudes and uncertainties on the 3D matter power spectrum.

Our work provides a qualitative consistency test of the $\Lambda$CDM model. Although we do not perform any thorough quantitative tests, we have computed the $\chi^2$ of the the data points shown in Fig.~\ref{fig:tegfig} against our fiducial model, ignoring any covariance between the data points, and using only the error bars in the $y$-direction. We find $\chi^2\,{=}\,117.3$ for 108 degrees of freedom, which is consistent with an expected $\chi^2$ fluctuation to within $1\,\sigma$. We stress that this number is only a very rough quantitative estimate of the consistency, but does at least highlight that no discrepancy is hiding in the residuals of Fig.~\ref{fig:tegfig}. Our results thus highlight the good agreement of the $\Lambda$CDM model with observational data issued by independent experiments, covering a large range of cosmic times and cosmic scales.

\section*{Acknowledgements}
We thank Guillaume Mention for valuable advice on the Total Variation Regularization technique, and Martin White and Benjamin Wallisch for helpful input. 

\bibliographystyle{mnras}
\bibliography{biblio,marius}

\let\jnlstyle=\rm\def\jref#1{{\jnlstyle#1}}\def\aj{\jref{AJ}}
  \def\araa{\jref{ARA\&A}} \def\apj{\jref{ApJ}\ } \def\apjl{\jref{ApJ}\ }
  \def\apjs{\jref{ApJS}} \def\ao{\jref{Appl.~Opt.}} \def\apss{\jref{Ap\&SS}}
  \def\aap{\jref{A\&A}} \def\aapr{\jref{A\&A~Rev.}} \def\aaps{\jref{A\&AS}}
  \def\azh{\jref{AZh}} \def\baas{\jref{BAAS}} \def\jrasc{\jref{JRASC}}
  \def\memras{\jref{MmRAS}} \def\mnras{\jref{MNRAS}\ }
  \def\pra{\jref{Phys.~Rev.~A}\ } \def\prb{\jref{Phys.~Rev.~B}\ }
  \def\prc{\jref{Phys.~Rev.~C}\ } \def\prd{\jref{Phys.~Rev.~D}\ }
  \def\pre{\jref{Phys.~Rev.~E}} \def\prl{\jref{Phys.~Rev.~Lett.}}
  \def\pasp{\jref{PASP}} \def\pasj{\jref{PASJ}} \def\qjras{\jref{QJRAS}}
  \def\skytel{\jref{S\&T}} \def\solphys{\jref{Sol.~Phys.}}
  \def\sovast{\jref{Soviet~Ast.}} \def\ssr{\jref{Space~Sci.~Rev.}}
  \def\zap{\jref{ZAp}} \def\nat{\jref{Nature}\ } \def\iaucirc{\jref{IAU~Circ.}}
  \def\aplett{\jref{Astrophys.~Lett.}}
  \def\apspr{\jref{Astrophys.~Space~Phys.~Res.}}
  \def\bain{\jref{Bull.~Astron.~Inst.~Netherlands}}
  \def\fcp{\jref{Fund.~Cosmic~Phys.}} \def\gca{\jref{Geochim.~Cosmochim.~Acta}}
  \def\grl{\jref{Geophys.~Res.~Lett.}} \def\jcp{\jref{J.~Chem.~Phys.}}
  \def\jgr{\jref{J.~Geophys.~Res.}}
  \def\jqsrt{\jref{J.~Quant.~Spec.~Radiat.~Transf.}}
  \def\memsai{\jref{Mem.~Soc.~Astron.~Italiana}}
  \def\nphysa{\jref{Nucl.~Phys.~A}} \def\physrep{\jref{Phys.~Rep.}}
  \def\physscr{\jref{Phys.~Scr}} \def\planss{\jref{Planet.~Space~Sci.}}
  \def\procspie{\jref{Proc.~SPIE}} \let\astap=\aap \let\apjlett=\apjl
  \let\apjsupp=\apjs \let\applopt=\ao \def\jcap{\jref{JCAP}}
  \let\jnlstyle=\rm\def\jref#1{{\jnlstyle#1}}\def\aj{\jref{AJ}}
  \def\araa{\jref{ARA\&A}} \def\apj{\jref{ApJ}\ } \def\apjl{\jref{ApJ}\ }
  \def\apjs{\jref{ApJS}} \def\ao{\jref{Appl.~Opt.}} \def\apss{\jref{Ap\&SS}}
  \def\aap{\jref{A\&A}} \def\aapr{\jref{A\&A~Rev.}} \def\aaps{\jref{A\&AS}}
  \def\azh{\jref{AZh}} \def\baas{\jref{BAAS}} \def\jrasc{\jref{JRASC}}
  \def\memras{\jref{MmRAS}} \def\mnras{\jref{MNRAS}\ }
  \def\pra{\jref{Phys.~Rev.~A}\ } \def\prb{\jref{Phys.~Rev.~B}\ }
  \def\prc{\jref{Phys.~Rev.~C}\ } \def\prd{\jref{Phys.~Rev.~D}\ }
  \def\pre{\jref{Phys.~Rev.~E}} \def\prl{\jref{Phys.~Rev.~Lett.}}
  \def\pasp{\jref{PASP}} \def\pasj{\jref{PASJ}} \def\qjras{\jref{QJRAS}}
  \def\skytel{\jref{S\&T}} \def\solphys{\jref{Sol.~Phys.}}
  \def\sovast{\jref{Soviet~Ast.}} \def\ssr{\jref{Space~Sci.~Rev.}}
  \def\zap{\jref{ZAp}} \def\nat{\jref{Nature}\ } \def\iaucirc{\jref{IAU~Circ.}}
  \def\aplett{\jref{Astrophys.~Lett.}}
  \def\apspr{\jref{Astrophys.~Space~Phys.~Res.}}
  \def\bain{\jref{Bull.~Astron.~Inst.~Netherlands}}
  \def\fcp{\jref{Fund.~Cosmic~Phys.}} \def\gca{\jref{Geochim.~Cosmochim.~Acta}}
  \def\grl{\jref{Geophys.~Res.~Lett.}} \def\jcp{\jref{J.~Chem.~Phys.}}
  \def\jgr{\jref{J.~Geophys.~Res.}}
  \def\jqsrt{\jref{J.~Quant.~Spec.~Radiat.~Transf.}}
  \def\memsai{\jref{Mem.~Soc.~Astron.~Italiana}}
  \def\nphysa{\jref{Nucl.~Phys.~A}} \def\physrep{\jref{Phys.~Rep.}}
  \def\physscr{\jref{Phys.~Scr}} \def\planss{\jref{Planet.~Space~Sci.}}
  \def\procspie{\jref{Proc.~SPIE}} \let\astap=\aap \let\apjlett=\apjl
  \let\apjsupp=\apjs \let\applopt=\ao \def\jcap{\jref{JCAP}}
  \let\jnlstyle=\rm\def\jref#1{{\jnlstyle#1}}\def\aj{\jref{AJ}}
  \def\araa{\jref{ARA\&A}} \def\apj{\jref{ApJ}\ } \def\apjl{\jref{ApJ}\ }
  \def\apjs{\jref{ApJS}} \def\ao{\jref{Appl.~Opt.}} \def\apss{\jref{Ap\&SS}}
  \def\aap{\jref{A\&A}} \def\aapr{\jref{A\&A~Rev.}} \def\aaps{\jref{A\&AS}}
  \def\azh{\jref{AZh}} \def\baas{\jref{BAAS}} \def\jrasc{\jref{JRASC}}
  \def\memras{\jref{MmRAS}} \def\mnras{\jref{MNRAS}\ }
  \def\pra{\jref{Phys.~Rev.~A}\ } \def\prb{\jref{Phys.~Rev.~B}\ }
  \def\prc{\jref{Phys.~Rev.~C}\ } \def\prd{\jref{Phys.~Rev.~D}\ }
  \def\pre{\jref{Phys.~Rev.~E}} \def\prl{\jref{Phys.~Rev.~Lett.}}
  \def\pasp{\jref{PASP}} \def\pasj{\jref{PASJ}} \def\qjras{\jref{QJRAS}}
  \def\skytel{\jref{S\&T}} \def\solphys{\jref{Sol.~Phys.}}
  \def\sovast{\jref{Soviet~Ast.}} \def\ssr{\jref{Space~Sci.~Rev.}}
  \def\zap{\jref{ZAp}} \def\nat{\jref{Nature}\ } \def\iaucirc{\jref{IAU~Circ.}}
  \def\aplett{\jref{Astrophys.~Lett.}}
  \def\apspr{\jref{Astrophys.~Space~Phys.~Res.}}
  \def\bain{\jref{Bull.~Astron.~Inst.~Netherlands}}
  \def\fcp{\jref{Fund.~Cosmic~Phys.}} \def\gca{\jref{Geochim.~Cosmochim.~Acta}}
  \def\grl{\jref{Geophys.~Res.~Lett.}} \def\jcp{\jref{J.~Chem.~Phys.}}
  \def\jgr{\jref{J.~Geophys.~Res.}}
  \def\jqsrt{\jref{J.~Quant.~Spec.~Radiat.~Transf.}}
  \def\memsai{\jref{Mem.~Soc.~Astron.~Italiana}}
  \def\nphysa{\jref{Nucl.~Phys.~A}} \def\physrep{\jref{Phys.~Rep.}}
  \def\physscr{\jref{Phys.~Scr}} \def\planss{\jref{Planet.~Space~Sci.}}
  \def\procspie{\jref{Proc.~SPIE}} \let\astap=\aap \let\apjlett=\apjl
  \let\apjsupp=\apjs \let\applopt=\ao \def\jcap{\jref{JCAP}}
  \let\jnlstyle=\rm\def\jref#1{{\jnlstyle#1}}\def\aj{\jref{AJ}}
  \def\araa{\jref{ARA\&A}} \def\apj{\jref{ApJ}\ } \def\apjl{\jref{ApJ}\ }
  \def\apjs{\jref{ApJS}} \def\ao{\jref{Appl.~Opt.}} \def\apss{\jref{Ap\&SS}}
  \def\aap{\jref{A\&A}} \def\aapr{\jref{A\&A~Rev.}} \def\aaps{\jref{A\&AS}}
  \def\azh{\jref{AZh}} \def\baas{\jref{BAAS}} \def\jrasc{\jref{JRASC}}
  \def\memras{\jref{MmRAS}} \def\mnras{\jref{MNRAS}\ }
  \def\pra{\jref{Phys.~Rev.~A}\ } \def\prb{\jref{Phys.~Rev.~B}\ }
  \def\prc{\jref{Phys.~Rev.~C}\ } \def\prd{\jref{Phys.~Rev.~D}\ }
  \def\pre{\jref{Phys.~Rev.~E}} \def\prl{\jref{Phys.~Rev.~Lett.}}
  \def\pasp{\jref{PASP}} \def\pasj{\jref{PASJ}} \def\qjras{\jref{QJRAS}}
  \def\skytel{\jref{S\&T}} \def\solphys{\jref{Sol.~Phys.}}
  \def\sovast{\jref{Soviet~Ast.}} \def\ssr{\jref{Space~Sci.~Rev.}}
  \def\zap{\jref{ZAp}} \def\nat{\jref{Nature}\ } \def\iaucirc{\jref{IAU~Circ.}}
  \def\aplett{\jref{Astrophys.~Lett.}}
  \def\apspr{\jref{Astrophys.~Space~Phys.~Res.}}
  \def\bain{\jref{Bull.~Astron.~Inst.~Netherlands}}
  \def\fcp{\jref{Fund.~Cosmic~Phys.}} \def\gca{\jref{Geochim.~Cosmochim.~Acta}}
  \def\grl{\jref{Geophys.~Res.~Lett.}} \def\jcp{\jref{J.~Chem.~Phys.}}
  \def\jgr{\jref{J.~Geophys.~Res.}}
  \def\jqsrt{\jref{J.~Quant.~Spec.~Radiat.~Transf.}}
  \def\memsai{\jref{Mem.~Soc.~Astron.~Italiana}}
  \def\nphysa{\jref{Nucl.~Phys.~A}} \def\physrep{\jref{Phys.~Rep.}}
  \def\physscr{\jref{Phys.~Scr}} \def\planss{\jref{Planet.~Space~Sci.}}
  \def\procspie{\jref{Proc.~SPIE}} \let\astap=\aap \let\apjlett=\apjl
  \let\apjsupp=\apjs \let\applopt=\ao \def\jcap{\jref{JCAP}}
\begin{thebibliography}{}
\makeatletter
\relax
\def\mn@urlcharsother{\let\do\@makeother \do\$\do\&\do\#\do\^\do\_\do\%\do\~}
\def\mn@doi{\begingroup\mn@urlcharsother \@ifnextchar [ {\mn@doi@}
  {\mn@doi@[]}}
\def\mn@doi@[#1]#2{\def\@tempa{#1}\ifx\@tempa\@empty \href
  {http://dx.doi.org/#2} {doi:#2}\else \href {http://dx.doi.org/#2} {#1}\fi
  \endgroup}
\def\mn@eprint#1#2{\mn@eprint@#1:#2::\@nil}
\def\mn@eprint@arXiv#1{\href {http://arxiv.org/abs/#1} {{\tt arXiv:#1}}}
\def\mn@eprint@dblp#1{\href {http://dblp.uni-trier.de/rec/bibtex/#1.xml}
  {dblp:#1}}
\def\mn@eprint@#1:#2:#3:#4\@nil{\def\@tempa {#1}\def\@tempb {#2}\def\@tempc
  {#3}\ifx \@tempc \@empty \let \@tempc \@tempb \let \@tempb \@tempa \fi \ifx
  \@tempb \@empty \def\@tempb {arXiv}\fi \@ifundefined
  {mn@eprint@\@tempb}{\@tempb:\@tempc}{\expandafter \expandafter \csname
  mn@eprint@\@tempb\endcsname \expandafter{\@tempc}}}

\bibitem[\protect\citeauthoryear{{Abolfathi} et~al.,}{{Abolfathi}
  et~al.}{2018}]{DR14eB}
{Abolfathi} B.,  et~al., 2018, \mn@doi [\apjs] {10.3847/1538-4365/aa9e8a},
  \href {http://adsabs.harvard.edu/abs/2018ApJS..235...42A} {235, 42}

\bibitem[\protect\citeauthoryear{{Armengaud}, {Palanque-Delabrouille},
  {Y{\`e}che}, {Marsh}  \& {Baur}}{{Armengaud} et~al.}{2017}]{Armengaud2017}
{Armengaud} E.,  {Palanque-Delabrouille} N.,  {Y{\`e}che} C.,  {Marsh}
  D.~J.~E.,   {Baur} J.,  2017, \mn@doi [\mnras] {10.1093/mnras/stx1870}, \href
  {http://adsabs.harvard.edu/abs/2017MNRAS.471.4606A} {471, 4606}

\bibitem[\protect\citeauthoryear{{Baur}, {Palanque-Delabrouille}, {Y{\`e}che},
  {Magneville}  \& {Viel}}{{Baur} et~al.}{2016}]{Baur2016}
{Baur} J.,  {Palanque-Delabrouille} N.,  {Y{\`e}che} C.,  {Magneville} C.,
  {Viel} M.,  2016, \mn@doi [\jcap] {10.1088/1475-7516/2016/08/012}, \href
  {http://adsabs.harvard.edu/abs/2016JCAP...08..012B} {8, 012}

\bibitem[\protect\citeauthoryear{{Baur}, {Palanque-Delabrouille}, {Y{\`e}che},
  {Boyarsky}, {Ruchayskiy}, {Armengaud}  \& {Lesgourgues}}{{Baur}
  et~al.}{2017}]{Baur2017}
{Baur} J.,  {Palanque-Delabrouille} N.,  {Y{\`e}che} C.,  {Boyarsky} A.,
  {Ruchayskiy} O.,  {Armengaud} {\'E}.,   {Lesgourgues} J.,  2017, \mn@doi
  [\jcap] {10.1088/1475-7516/2017/12/013}, \href
  {http://adsabs.harvard.edu/abs/2017JCAP...12..013B} {12, 013}

\bibitem[\protect\citeauthoryear{{Bautista} et~al.,}{{Bautista}
  et~al.}{2017}]{Bautista2017}
{Bautista} J.~E.,  et~al., 2017, \mn@doi [\aap] {10.1051/0004-6361/201730533},
  \href {http://adsabs.harvard.edu/abs/2017A%26A...603A..12B} {603, A12}

\bibitem[\protect\citeauthoryear{{Bennett} et~al.,}{{Bennett}
  et~al.}{2013}]{WMAP9yr}
{Bennett} C.~L.,  et~al., 2013, \mn@doi [\apjs] {10.1088/0067-0049/208/2/20},
  \href {https://ui.adsabs.harvard.edu/abs/2013ApJS..208...20B} {208, 20}

\bibitem[\protect\citeauthoryear{{Blanton} et~al.,}{{Blanton}
  et~al.}{2017}]{Blanton2017}
{Blanton} M.~R.,  et~al., 2017, \mn@doi [\aj] {10.3847/1538-3881/aa7567}, \href
  {http://adsabs.harvard.edu/abs/2017AJ....154...28B} {154, 28}

\bibitem[\protect\citeauthoryear{{Borde}, {Palanque-Delabrouille}, {Rossi},
  {Viel}, {Bolton}, {Y{\`e}che}, {LeGoff}  \& {Rich}}{{Borde}
  et~al.}{2014}]{Borde2014}
{Borde} A.,  {Palanque-Delabrouille} N.,  {Rossi} G.,  {Viel} M.,  {Bolton}
  J.~S.,  {Y{\`e}che} C.,  {LeGoff} J.-M.,   {Rich} J.,  2014, \mn@doi [\jcap]
  {10.1088/1475-7516/2014/07/005}, \href
  {http://adsabs.harvard.edu/abs/2014JCAP...07..005B} {7, 5}

\bibitem[\protect\citeauthoryear{Busca et~al.,}{Busca et~al.}{2013}]{Busca2012}
Busca N.~G.,  et~al., 2013, \mn@doi [Astronomy \& Astrophysics]
  {10.1051/0004-6361/201220724}, 552, A96

\bibitem[\protect\citeauthoryear{Chabanier, Palanque-Delabrouille, Y{\`{e}}che
  et~al.}{Chabanier et~al.}{2018}]{Chabanier2018}
Chabanier S.,  Palanque-Delabrouille N.,  Y{\`{e}}che C.,   et~al., 2018,
  arXiv:1812.03554 [astro-ph]

\bibitem[\protect\citeauthoryear{Chartrand}{Chartrand}{2005}]{Chartrand2005}
Chartrand R.,  2005, ISRN Applied Mathematics, 2011, Article ID 164564

\bibitem[\protect\citeauthoryear{Collaboration et~al.,}{Collaboration
  et~al.}{2018a}]{planck2018legacy}
Collaboration P.,  et~al., 2018a, arXiv:1807.06205 [astro-ph]

\bibitem[\protect\citeauthoryear{Collaboration et~al.,}{Collaboration
  et~al.}{2018b}]{planck2018lensing}
Collaboration P.,  et~al., 2018b, arXiv:1807.06210 [astro-ph]

\bibitem[\protect\citeauthoryear{{Colless} et~al.,}{{Colless}
  et~al.}{2001}]{2dFGRS}
{Colless} M.,  et~al., 2001, \mn@doi [\mnras]
  {10.1046/j.1365-8711.2001.04902.x}, \href
  {http://adsabs.harvard.edu/abs/2001MNRAS.328.1039C} {328, 1039}

\bibitem[\protect\citeauthoryear{Croft, Weinberg, Katz  \& Hernquist}{Croft
  et~al.}{1998}]{Croft1998}
Croft R. A.~C.,  Weinberg D.~H.,  Katz N.,   Hernquist L.,  1998, \mn@doi [The
  Astrophysical Journal] {10.1086/305289}, 495, 44

\bibitem[\protect\citeauthoryear{Croft, Weinberg, Bolte, Burles, Hernquist,
  Katz, Kirkman  \& Tytler}{Croft et~al.}{2002}]{Croft2002}
Croft R. A.~C.,  Weinberg D.~H.,  Bolte M.,  Burles S.,  Hernquist L.,  Katz
  N.,  Kirkman D.,   Tytler D.,  2002, \mn@doi [The Astrophysical Journal]
  {10.1086/344099}, 581, 20

\bibitem[\protect\citeauthoryear{{Dawson} et~al.,}{{Dawson}
  et~al.}{2016}]{Dawson2016}
{Dawson} K.~S.,  et~al., 2016, \mn@doi [\aj] {10.3847/0004-6256/151/2/44},
  \href {http://adsabs.harvard.edu/abs/2016AJ....151...44D} {151, 44}

\bibitem[\protect\citeauthoryear{{Delubac} et~al.,}{{Delubac}
  et~al.}{2015}]{Delubac2015}
{Delubac} T.,  et~al., 2015, \mn@doi [\aap] {10.1051/0004-6361/201423969},
  \href {http://adsabs.harvard.edu/abs/2015A%26A...574A..59D} {574, A59}

\bibitem[\protect\citeauthoryear{Hlozek et~al.,}{Hlozek
  et~al.}{2012}]{hlozek2012}
Hlozek R.,  et~al., 2012, \mn@doi [The Astrophysical Journal]
  {10.1088/0004-637X/749/1/90}, 749, 90

\bibitem[\protect\citeauthoryear{{Hoekstra}, {Yee}  \& {Gladders}}{{Hoekstra}
  et~al.}{2002}]{RSCS}
{Hoekstra} H.,  {Yee} H. K.~C.,   {Gladders} M.~D.,  2002, \mn@doi [\apj]
  {10.1086/342120}, \href
  {https://ui.adsabs.harvard.edu/abs/2002ApJ...577..595H} {577, 595}

\bibitem[\protect\citeauthoryear{{Ir{\v s}i{\v c}} et~al.,}{{Ir{\v s}i{\v c}}
  et~al.}{2017}]{Irsic2017}
{Ir{\v s}i{\v c}} V.,  et~al., 2017, \mn@doi [\prd]
  {10.1103/PhysRevD.96.023522}, \href
  {http://adsabs.harvard.edu/abs/2017PhRvD..96b3522I} {96, 023522}

\bibitem[\protect\citeauthoryear{{Kirkby} et~al.,}{{Kirkby}
  et~al.}{2013}]{Kirkby2013}
{Kirkby} D.,  et~al., 2013, \mn@doi [\jcap] {10.1088/1475-7516/2013/03/024},
  \href {http://adsabs.harvard.edu/abs/2013JCAP...03..024K} {3, 024}

\bibitem[\protect\citeauthoryear{{Lesgourgues}}{{Lesgourgues}}{2011}]{CLASS}
{Lesgourgues} J.,  2011, preprint, \href
  {http://adsabs.harvard.edu/abs/2011arXiv1104.2932L} {} (\mn@eprint {arXiv}
  {1104.2932})

\bibitem[\protect\citeauthoryear{Lewis, Challinor  \& Lasenby}{Lewis
  et~al.}{2000}]{Lewis2000}
Lewis A.,  Challinor A.,   Lasenby A.,  2000, \mn@doi [The Astrophysical
  Journal] {10.1086/309179}, 538, 473

\bibitem[\protect\citeauthoryear{McDonald et~al.,}{McDonald
  et~al.}{2006}]{McDonald2006}
McDonald P.,  et~al., 2006, \mn@doi [The Astrophysical Journal Supplement
  Series] {10.1086/444361}, 163, 80

\bibitem[\protect\citeauthoryear{{Palanque-Delabrouille}
  et~al.,}{{Palanque-Delabrouille} et~al.}{2015a}]{Palanque2015a}
{Palanque-Delabrouille} N.,  et~al., 2015a, \mn@doi [\jcap]
  {10.1088/1475-7516/2015/02/045}, \href
  {http://adsabs.harvard.edu/abs/2015JCAP...02..045P} {2, 45}

\bibitem[\protect\citeauthoryear{{Palanque-Delabrouille}
  et~al.,}{{Palanque-Delabrouille} et~al.}{2015b}]{Palanque2015b}
{Palanque-Delabrouille} N.,  et~al., 2015b, \mn@doi [\jcap]
  {10.1088/1475-7516/2015/11/011}, \href
  {http://adsabs.harvard.edu/abs/2015JCAP...11..011P} {11, 11}

\bibitem[\protect\citeauthoryear{{Planck Collaboration} et~al.,}{{Planck
  Collaboration} et~al.}{2014}]{PlanckCollaboration2013}
{Planck Collaboration} et~al., 2014, \mn@doi [\aap]
  {10.1051/0004-6361/201321591}, \href
  {http://adsabs.harvard.edu/abs/2014A%26A...571A..16P} {571, A16}

\bibitem[\protect\citeauthoryear{{Planck Collaboration} et~al.,}{{Planck
  Collaboration} et~al.}{2018}]{PlanckCosmo2018}
{Planck Collaboration} et~al., 2018, arXiv:1807.06209 [astro-ph], \href
  {http://adsabs.harvard.edu/abs/2018arXiv180706209P} {}

\bibitem[\protect\citeauthoryear{Reid et~al.,}{Reid et~al.}{2010}]{reid2010a}
Reid B.~A.,  et~al., 2010, \mn@doi [Monthly Notices of the Royal Astronomical
  Society] {10.1111/j.1365-2966.2010.16276.x}, 404, 60

\bibitem[\protect\citeauthoryear{{Slosar} et~al.,}{{Slosar}
  et~al.}{2011}]{Slosar2011}
{Slosar} A.,  et~al., 2011, \mn@doi [\jcap] {10.1088/1475-7516/2011/09/001},
  \href {http://adsabs.harvard.edu/abs/2011JCAP...09..001S} {9, 001}

\bibitem[\protect\citeauthoryear{Slosar et~al.,}{Slosar
  et~al.}{2013}]{Slosar2013}
Slosar A.,  et~al., 2013, \mn@doi [Journal of Cosmology and Astroparticle
  Physics] {10.1088/1475-7516/2013/04/026}, 2013, 026

\bibitem[\protect\citeauthoryear{Tegmark \& Zaldarriaga}{Tegmark \&
  Zaldarriaga}{2002}]{Tegmark2002}
Tegmark M.,  Zaldarriaga M.,  2002, \mn@doi [Physical Review D]
  {10.1103/PhysRevD.66.103508}, 66, 103508

\bibitem[\protect\citeauthoryear{Tegmark \& Zaldarriaga}{Tegmark \&
  Zaldarriaga}{2009}]{tegmark2009}
Tegmark M.,  Zaldarriaga M.,  2009, \mn@doi [Physical Review D]
  {10.1103/PhysRevD.79.083530}, 79, 083530

\bibitem[\protect\citeauthoryear{{The Dark Energy Survey Collaboration}}{{The
  Dark Energy Survey Collaboration}}{2005}]{DES}
{The Dark Energy Survey Collaboration} 2005, arXiv e-prints, \href
  {https://ui.adsabs.harvard.edu/abs/2005astro.ph.10346T} {pp
  astro--ph/0510346}

\bibitem[\protect\citeauthoryear{Troxel et~al.,}{Troxel
  et~al.}{2017}]{troxel2017}
Troxel M.~A.,  et~al., 2017, arXiv:1708.01538 [astro-ph]

\bibitem[\protect\citeauthoryear{{Y{\`e}che}, {Palanque-Delabrouille}, {Baur}
  \& {du Mas des Bourboux}}{{Y{\`e}che} et~al.}{2017}]{Yeche2017}
{Y{\`e}che} C.,  {Palanque-Delabrouille} N.,  {Baur} J.,   {du Mas des
  Bourboux} H.,  2017, \mn@doi [\jcap] {10.1088/1475-7516/2017/06/047}, \href
  {http://adsabs.harvard.edu/abs/2017JCAP...06..047Y} {6, 047}

\bibitem[\protect\citeauthoryear{York et~al.,}{York et~al.}{2000}]{York2000}
York D.~G.,  et~al., 2000, \mn@doi [The Astronomical Journal] {10.1086/301513},
  120, 1579

\bibitem[\protect\citeauthoryear{Zaldarriaga}{Zaldarriaga}{1997}]{zaldarriaga1997a}
Zaldarriaga M.,  1997, \mn@doi [Physical Review D] {10.1103/PhysRevD.55.1822},
  55, 1822

\makeatother
\end{thebibliography}

% Don't change these lines
\bsp	% typesetting comment
\label{lastpage}
\end{document}